\documentclass[prl,preprint]{revtex4-1}
\usepackage{times}
\usepackage{braket}
\usepackage{amsmath}
\usepackage{graphicx}

\begin{document}

\title{Conditional preparation of non-Gaussian quantum optical states by mesoscopic measurement}


	\author{Alex O. C. Davis$^{1,2}$, Mattia Walschaers$^{1}$, Valentina Parigi$^{1}$, Nicolas Treps$^{1}$}


\address{{$^{1}$}Laboratoire Kastler Brossel, Sorbonne Universit\'{e}, CNRS, ENS-Universit\'{e} PSL, Coll\`{e}ge de France; 4 place Jussieu, F-75252 Paris, France}
\address{{$^{2}$}Centre for Photonics and Photonic Materials, Department of Physics, University of Bath, Bath BA2 7AY, UK}

\date{\today}

\begin{abstract}
Non-Gaussian states of an optical field are important as a proposed resource in quantum information applications. While conditional preparation is a highly successful approach to preparing such states, their quality is limited by detector non-idealities such as dead time, narrow dynamic range, limited quantum efficiency and dark noise. Mesoscopic photon counters, with peak performance at higher photon number, offer many practical advantages over single-photon level conditioning detectors. Here we propose a novel approach involving displacement of the ancilla field into the regime where mesoscopic detectors can be used.  We explore this strategy theoretically and present simulations accounting for experimental non-idealities such as loss and amplification noise, showing that precise photon-number resolution is not necessary to herald highly nonclassical states. We conclude that states with strong Wigner negativity can be prepared at high rates by this technique under experimentally attainable conditions.
\end{abstract}


\maketitle

\section{Introduction}

Quantum information processing (QIP) on optical platforms with better scaling than can be achieved classically is known to require states with non-positive Wigner functions \cite{Mari2012}. In the ideal case of pure states, a sufficient condition for this is if the wave function is described by a non-Gaussian function of the field quadratures \cite{Hudson1974}. The experimental preparation of such non-Gaussian states is therefore of paramount interest across the quantum optics and quantum information communities. 

Strategies for preparing such exotic quantum states of light can be classified into two groups. The first of these categories is \emph{deterministic} preparation, where non-Gaussianity is induced unitarily by a Hamiltonian that is nonlinear in the field. This class encompasses ``push-button'' single-photon sources based on interactions in a quantum dot, single atom or crystal defect \cite{senellart2017high}. Despite recent progress in the technical development of these sources, they can suffer from adverse properties including poor indistinguishability, narrow bandwidth and low mode number. Meanwhile, attempts to prepare non-Gaussian states deterministically using bulk nonlinearities (e.g. by photon triplet generation in $\chi_3$ media \cite{corona2011experimental}) have not yet succeeded in showing Wigner negativity. 

The second approach is \emph{conditional} non-Gaussian state preparation. A multimode state exhibiting Gaussian entanglement is prepared unitarily, typically by parametric down conversion (PDC) or four-wave mixing. One mode, known as the ancilla, is then subjected to a \emph{conditioning} measurement in a non-Gaussian basis, often the Fock basis, collapsing the wave function over the remaining mode(s), the \emph{signal}, into a non-Gaussian state \cite{Quesada2019}. The outcome of the conditioning measurement must be recorded and used in conjunction with any application of the signal: if this information is discarded, the signal is described by an impure Gaussian state and hence is less useful for QIP. The most widely-used technique in this class is heralded single photon preparation \cite{mosley2008heralded}, although it has also been used in the preparation of higher-order Fock states \cite{cooper2013experimental}, photon-subtracted squeezed states \cite{Ra2019}, and many other schemes.

 Conditionally prepared states can have many advantages such as broad spectral bandwidth, good indistinguishability, high collection efficiency, high purity and multimode structure. However, they suffer a fundamental drawback in that conditional preparation is intrinsically probabilistic: projective measurements of just one part of an entangled state will always have many possible outcomes and so the final state impossible to know \emph{a priori}. Additionally, for conditioning with direct photon counting, the vacuum is itself a Gaussian state so non-detection events, which are the most likely measurement outcomes in the case of low photon number, cannot be used to induce non-Gaussianity.

Conditional preparation is also at the mercy of the physical properties of the conditioning detector. At the single-photon level, these often deviate significantly from the ideal \cite{Hadfield2009}. Commercially available single-photon avalanche detectors (SPADs) and superconducting nanowire single-photon detectors (SNSPDs) typically saturate at the single-photon level, rendering them unable to distinguish between single-photon and multi-photon events and hence reducing the purity of states heralded by such events. While recent advances in SNSPD technology and the technique of pseudo-number resolved detection propose to mitigate this, these remain at an early level of development or introduce additional experimental challenges \cite{zhu2020resolving,Zhai:13}. Many such detectors also have a characteristic \emph{dead time}, during which the detector is blind to new events for some period after a detection. These effects often mean that the average photon number in the conditioning mode must be kept well below one, to achieve purity at the price of making non-Gaussian state preparation highly inefficient. Photon counting technologies such as transition edge sensors exhibit low bandwidth/timing resolution, which sets a practical limitation on the rate at which they can be used to prepare non-Gaussian states. Other considerations include the quantum efficiency (which impacts the ability to distinguish between nearby Fock states and hence the purity), dark counts, and price, since some technologies operate at low temperature and hence must be situated in expensive refrigerators.

An emerging class of quantum receiver is the mesoscopic detectors (MSDs), which are used to resolve photon number in a regime intermediate between the single-photon level and the classical scale. Although generally too noisy for true number-resolved detection, MSDs exhibit the sub-Poissonian number resolution necessary for inducing non-Gaussianity and a combination of other desirable properties that at present cannot be found among single-photon level detectors. Much progress has been made in recent developments of this class of detectors, with multiple design paradigms including silicon photomultipliers \cite{Hesi2019}, avalanche photodiode devices \cite{vojetta2012linear, Dumas2017,sun2014hgcdte} and, most recently, superconducting nanostrips \cite{endo2021}. Potential advantages of MSDs over single photon counters include cost-effectiveness and easier temperature control, fast response, wide dynamic range, no meaningful dead time and high quantum efficiency. Additionally, operating in the mesoscopic regime implies a low probability of non-detection events, so some non-Gaussianity is induced for all the most likely measurement outcomes. Together, these properties create the potential for rapid preparation of broadband, high-purity non-Gaussian states with easy generalisation to multimode operation. 

Previous work has shown the usefulness of photon counting at mesoscopic scales for QIP applications. For example, by splitting single-mode squeezed vacuum at a 50:50 beam splitter and conditioning off high Fock state measurements of one output beam, the states generated  are approximately Schr\"{o}dinger cat states \cite{Dakna97}. Schr\"{o}dinger cat states are known to be useful as a resource for generating Gottesman-Kitaev-Preskill states \cite{eaton2019non} and hence universal error-corrected quantum computing \cite{Menicucci:14}. The necessary conditioning measurement for this scheme would be in the mesoscopic regime for realistic squeezing levels. However, preparing Schr\"{o}dinger cat states in this way is not a practical approach to QIP, since the heralding events are rare and the nonclassicality of the post-selected states is not robust even to  single-photon errors in counting.

Here, we propose a practical conditional preparation technique that circumvents the problems of using single photon counters whilst still allowing the rapid conditional preparation of non-Gaussian states with realistic detectors. This is achieved by displacing the ancilla and performing the conditioning detection at intermediate photon numbers. We show that displacing the ancilla prior to photon counting makes the Wigner negativity of the output states robust to errors in photon counting comparable to those encountered with currently achievable technology. By combining displaced Fock state measurement with mesoscopic detectors, we therefore show that it is feasible to generate multi-photon states exhibiting strong Wigner negativity under realistic experimental conditions with relatively high rates, exceeding that of heralded single photon state preparation using SPADs.

\begin{figure}[t!]
\includegraphics[width=\linewidth]{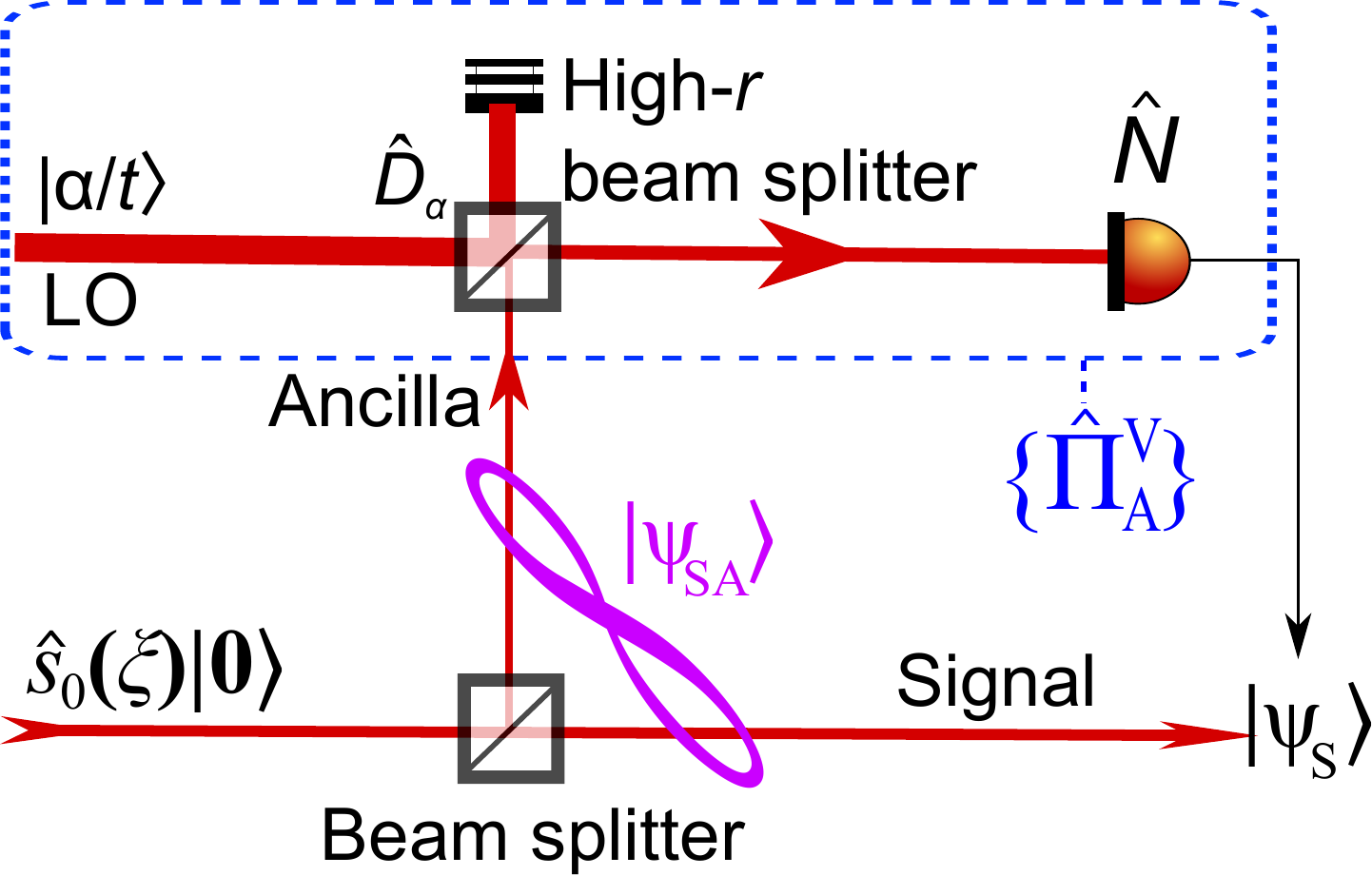}
\caption{Schematic diagram of the ideal mesoscopic conditioning scheme. A single-mode squeezed vacuum state is split at a 50:50 beam splitter into a signal and ancilla beam, which together constitute the state $\Ket{\psi}_{SA}$. The ancilla is displaced by interference with a bright coherent state (LO- local oscillator) at a high reflectivity beam coupler and undergoes measurement in a mesoscopic photon counter. The output of the measurement is used to herald the signal state $\Ket{\psi}_S$. }
\label{schematic}
\end{figure}

\section{Ideal displaced Fock state measurement}
\subsection{Ideal-case POVMs}
In the general single-mode conditional measurement scheme, a two-mode Gaussian state $\hat{\rho}_{SA}$ with entanglement between the signal $S$ and ancilla $A$ modes is prepared unitarily, and the ancilla is subjected to a conditioning measurement described by a positive operator-valued measure (POVM) with elements $\{\hat{\Pi}_A^V\}$ and possible outcomes indexed $V$ (which may correspond to the voltage output of the detector, for example). The post-measurement state of the signal after an outcome $V$  is given by

\begin{equation}
\hat{\rho}_S^V=\frac{1}{N}\mbox{Tr}_A\{\hat{\rho}_{SA}\hat{\Pi}_A^V\}.
\end{equation}

where $\mbox{Tr}_A$ indicates the trace over the ancilla subsystem and $N$ is a normalisation constant. Here, we consider that the ideal measurement operation consists of displacement followed by a Fock state measurement. In this case, the measurement outcomes $V$ correspond precisely to photon number states on the ancilla $\Ket{n}_A$ and so the POVM elements are given by

\begin{equation}
\hat{\Pi}_A^n= \hat{D}_A(\alpha)\Ket{n}_A\Bra{n}\hat{D}^\dagger_A(\alpha),
\label{projective}
\end{equation}

where $\Ket{n}_A$ is the $n$-th Fock state and $\hat{D}_A(\alpha)$ is the displacement operator:

\begin{equation}
\hat{D}_A(\alpha)=\exp\{\alpha \hat{a}_A^\dagger-\alpha^*\hat{a}_A\},
\end{equation}

where $\alpha$ is the displacement parameter and $\hat{a}_A$ and $\hat{a}_A^\dagger$ the bosonic creation and annihilation operators for the mode $A$, respectively. For a given value of $\alpha$ the basis of displaced Fock states $\{\hat{D}_A(\alpha)\Ket{n}_A\}$, is a complete orthonormal basis over the field and hence $\{\hat{\Pi}_A^n\}$ constitutes a projective measurement.

Experimentally, it is straightforward to approximate the action of the displacement operator by reflecting the beam off a beam coupler with near-unit reflectivity $r$ that is back-lit by a relatively bright coherent state with displacement parameter $\alpha/\sqrt{1-r^2}$. This procedure corresponds to ideal displacement in the limit $r\rightarrow1$ \cite{Paris:96}. The measure $\{\hat{\Pi}_A^n\}$ can therefore be realised by such displacement followed by intensity (photon-number) resolving detection, with $\alpha$ chosen such that the mean photon number after displacement is in the mesoscopic regime.

By attributing the eigenvalue $n$ to the $n$-th measurement outcome, we define the displaced Fock state observable

\begin{align}
\hat{\Upsilon}_A(\alpha)\equiv & \sum_n n\hat{\Pi}_A^n=\sum_n n\hat{D}_A(\alpha)\Ket{n}_A\Bra{n}\hat{D}^\dagger_A(\alpha) \\
=& \hat{D}_A(\alpha)\hat{a}_A^\dagger \hat{a}_A\hat{D}^\dagger_A(\alpha). \nonumber
\end{align}

Employing the commutation relation 

\begin{equation}
\left[ \hat{D}_A(\alpha),\hat{a}_A^\dagger \right]=\alpha \hat{D}_A(\alpha),
\end{equation}

we can rewrite $\hat{\Upsilon}_A(\alpha)$ as

\begin{equation}
\hat{\Upsilon}_A(\alpha)= |\alpha|^2 + (\alpha \hat{a}_A^\dagger+\alpha^*\hat{a}_A) + \hat{a}_A^\dagger \hat{a}_A.
\end{equation}

This operator is a sum of three terms: a constant term $|\alpha|^2$ (which only affects the first moment of $\hat{\Upsilon}_A(\alpha)$), a ``homodyne'' term $(\alpha \hat{a}_A^\dagger+\alpha^*\hat{a}_A)$ and a ``photon-counting'' term $\hat{a}_A^\dagger \hat{a}_A$. In the case where $\alpha \rightarrow 0$, $\hat{\Upsilon}_A(0)$ simply reduces to the bare photon-counting operator $\hat{a}_A^\dagger \hat{a}_A$. In the other limit $\alpha \rightarrow \infty$, the moments of the homodyne term dominate and $\hat{\Upsilon}_A(\alpha)$ reduces to the unbalanced homodyne detection operator \cite{wallentowitz1996unbalanced}, which is a Gaussian operation and so preserves the Gaussianity of the state. 

We therefore note that there is a trade off, tuned by the displacement parameter $\alpha$, between strong non-Gaussianity at low $\alpha$, necessitating single-photon level detection, and weak non-Gaussianity at high $\alpha$, in the many-photon regime where detectors have more desirable properties. The desire to optimise this trade-off motivates our exploration of the mesoscopic regime.


\subsection{Conditioning a single-mode squeezed state split at a beam splitter}
We now consider the form of the states produced in the ideal case with a particular class of input state. A schematic for this is shown in Fig. \ref{schematic}. We assume that the Gaussian-entangled input state is the pure state generated by shining a single-mode squeezed vacuum state in the mode $\hat{a}_0$ onto a 50:50 beam splitter. We can represent this state as $\hat{\rho}_{SA}=\Ket{\psi}_{SA}\Bra{\psi}$, where

\begin{align}
\Ket{\psi}_{SA}=&\hat{s}_0(\xi)\Ket{\mbox{vac}} \\
=&\sum_{j=0}^\infty c'_j(a_0^\dagger)^{2j} \Ket{\mbox{vac}} \\
=&\sum_{j=0}^\infty c_j(\hat{a}_A^\dagger+\hat{a}_S^\dagger)^{2j} \Ket{\mbox{vac}} \label{xi}
\end{align}

where $c'_j=2^jc_j$, $\hat{s}_0(\xi) \equiv \exp\{\xi \hat{a}_0^2 +\xi^* \hat{a}_0^{\dagger 2}\}$ is the squeezing operator, $\xi$ is the squeezing parameter and $\Ket{\mbox{vac}}$ is the global vacuum state, and
\begin{equation}
c_j=\frac{1}{\sqrt{\cosh r}} \frac{(-1)^j}{4^jj!}(e^{i\theta}\tanh r)^j
\end{equation}
with $\xi=re^{i\theta}$ and $r$ real and non-negative. 

For this ``split squeezed vacuum state'',  the reduced states on either mode yielded by tracing over the other have an intrinsic phase sensitivity, with local quadrature noise variation on each mode defining a local phase determined by $\theta$, which we will from now on set $\theta=0$ without loss of generality. The presence of a physically significant local phase on the ancilla means that the relative phase of the displacement $\alpha$ has an effect on the form of the conditioning operation. In particular, by controlling this phase we can choose to displace the ancilla along either the squeezed or anti-squeezed quadratures of the ancilla field prior to measurement. As we will show in our simulations, when displacing along the anti-squeezed quadrature, the homodyne term of the measurement operator, $\alpha \hat{a}_A^\dagger+\alpha^*\hat{a}_A$, has large higher moments and dominates the statistics of $\hat{\Upsilon}_A$. The conditional projection is therefore more similar to a (Gaussian) homodyne measurement. By contrast, when the displacement is along the squeezed quadrature, the photon counting term contributes more to the measurement outcome distribution, leading to a higher degree of induced non-Gaussianity.

We now calculate the explicit form of the states conditioned by MSDs. Using the binomial theorem and re-ordering the summation we write Eq. \ref{xi} as

\begin{align}
\ket{\psi}_{SA} = & \sum_{j=0}^\infty c_j \left[ \sum_{k=0}^{2j} \frac{(2j)!}{(2j-k)!k!} \hat{a}_A^{\dagger k}\hat{a}_S^{\dagger(2j-k)} \right] \Ket{\mbox{vac}} \\
 = & \sum_{k=0}^{\infty}\left[\sum_{j=\lceil\frac{k}{2}\rceil}^\infty c_j \frac{(2j)!}{(2j-k)!k!} \hat{a}_S^{\dagger(2j-k)} \right] \hat{a}_A^{\dagger k}\Ket{\mbox{vac}} \\
= & \sum_{k=0}^{\infty}\left[\sum_{j=\lceil\frac{k}{2}\rceil}^\infty c_j \frac{(2j)!}{(2j-k)!\sqrt{k!}} \hat{a}_S^{\dagger(2j-k)} \right] \Ket{0}_S\Ket{k}_A,
\end{align}
where $\lceil.\rceil$ is the ceiling function. The state on the signal mode prepared by measurement of the displaced Fock state $\hat{D}_A^\dagger(\alpha)\Ket{m}$ in the ancilla mode is then
\begin{align}
\Ket{\psi_m}_S=&N_m\Bra{m}\hat{D}_A(\alpha)\Ket{\psi}_{SA} \\
=& N_m\sum_{k=0}^{\infty}\left[\sum_{j=\lceil\frac{k}{2}\rceil}^\infty c_j \frac{(2j)!}{(2j-k)!\sqrt{k!}} \hat{a}_S^{\dagger(2j-k)} \right] \Delta_{mk} \Ket{0}_S , \label{xip}
\end{align}
where $N_m$ is a normalisation constant and we have defined $\Delta_{mk}\equiv \Bra{m}\hat{D}(\alpha)\Ket{k}$, the matrix elements of the displacement operator in the Fock basis. Written explicitly,

\begin{equation}
\Delta_{mk}=e^{-\frac{|\alpha|^2}{2}}(-\alpha^*)^k\alpha^m \sum_{n=0}^{\text{min}(m,k)}\frac{\sqrt{k!m!}(-1)^n|\alpha|^{-2n}}{n!(k-n)!(m-n!)}.
\label{deltapk}
\end{equation}
Note that since the displacement operator tends to the identity for weak displacements, $\Delta_{mk} \rightarrow \delta_{mk}$ as $|\alpha|^2 \ll \{m,k\}$.

It is illuminating to consider the terms of opposite parity in the heralded state $\Ket{\psi_m}_S$. Dividing Eq. \ref{xip} into the terms where $k$ is either even or odd 
and rearranging the series, we obtain the final form for the pure states heralded by diplaced Fock state detection,
\begin{align}
\Ket{\psi_m}_S=&N_m\sum_{j=0}^\infty  \left( \left[ \sum_{k \text{ even}}^{\infty} c_{j+\frac{k}{2}} \frac{(2j+k)!}{\sqrt{(2j)!k!}} \Delta_{mk} \right] \right. \Ket{2j}_S  \label{parity}\\ 
+& \left[ \left. \sum_{k \text{ odd}}^{\infty} c_{j+\frac{k+1}{2}} \frac{(2j+k+1)!}{\sqrt{(2j+1)!k!}}  \Delta_{mk} \right]\right)  \Ket{2j+1}_S . \nonumber
\end{align}
The dependence of these two collections of terms on the phase of the displacement matrix elements $\Delta_{mk}$ turns out to have an important role in the degree to which $\Ket{\psi_m}_S$ can be well approximated by a Gaussian state.

\section{Role of displacement phase}

\label{phase}

In the Appendix we show that in the zero-displacement case (where $\Delta_{mk}=\delta_{mk}$) equation \ref{parity} tends towards a Schr\"{o}dinger cat state as $m \rightarrow \infty$, recovering the result found in \cite{Dakna97}.

We now consider a small nonzero displacement, nonetheless introducing fewer photons than the conditioning count (i.e. $|\alpha|^2 \ll m$).  Therefore, the final term in the sum in Eq. \ref{deltapk} dominates. In this case, the displacement is small enough that the state $\hat{D}_A^\dagger(\alpha)\Ket{m}$ has contributions from only a few Fock states $\Ket{k}$ centred on $k=m$. Starting from the final (dominant) term in the sum in Eq. \ref{deltapk} $(n=m)$, and taking $\delta\equiv k-m$ we can write
\begin{equation}
\Delta_{m,m+\delta}=e^{-\frac{|\alpha|^2}{2}}(-1)^{\delta}(\alpha^*)^\delta\frac{\sqrt{(m+\delta)!m!}}{m!\delta!}.
\end{equation}
With the approximation $(m+\delta)!\approx m^\delta m!$ for $m \gg \delta$, this reduces to
\begin{equation}
\Delta_{m,(m+\delta)} \approx e^{-|\alpha|^2} \frac{m^\frac{|\delta|}{2}}{|\delta|!}(-1)^{\delta}|\alpha|^\delta e^{-i\phi\delta},
\label{neardiagonals}
\end{equation} 
where $\alpha=|\alpha|e^{i\phi}$. Note the dependence of the phase of $\Delta_{m,(m+\delta)}$ on the term $e^{-i\phi\delta}$.

We are now in a position to understand the role of the phase of $\alpha$. Consider the terms in Eq. \ref{parity} where $m$ and $k$ have opposite parity to $m$, such that $\delta=\{\pm1,\pm3,\pm5...\}$. If $\alpha$ is real (i.e. $\phi=\{0,\pi\}$) then the two matrix elements of the same order in $\alpha$, $\Delta_{m,(m+\delta)}$ and $\Delta_{m,(m-\delta)}$, have the same sign. Because $c_{m+k/2}$ and $c_{m+(k+1)/2}$ both have opposite sign for $\pm \delta$, the terms at $\pm \delta$ in the sum over $k$ add together destructively and the contribution of Fock states with parity opposite to $m$ is suppressed. By contrast, the terms with the same parity have $\delta=\{0,\pm2,\pm4...\}$ and the terms in the sum over $k$ have the same sign regardless of whether $\alpha$ is real or purely imaginary. Hence terms with the same parity as $m$ are reinforced. In this case, the state once again resembles a Schr\"{o}dinger cat state.

By contrast, if $\alpha$ is purely imaginary (i.e. $\phi=\{\pi/2,3\pi/2\}$) then the terms in both the even and odd-parity series sum constructively. In this case, once $\alpha$ is large enough that $\hat{D}_A^\dagger(\alpha)\Ket{m}$ contains large contributions from several Fock states adjacent to $\Ket{m}$, the contributions of both the even-parity and odd-parity series become comparable. Following similar reasoning to that in Section \ref{focklimit}, the state then approximates a coherent state, which is of limited interest for QIP.

Whilst Schr\"{o}dinger cat states are undoubtedly useful as a nonclassical resource, unfortunately this limit (where the conditioning count number $m$ is large relative to $|\alpha|^2$) comes with two disadvantages. Firstly, detecting $m$ photons in the ancilla arm is unlikely for large $m$, meaning preparation of these states is inefficient. Secondly, the nonclassicality of the signal derives from its strongly nonzero parity, which is determined by the parity of $m$. The conditioning measurement must therefore be precise enough to accurately determine the parity the ancilla, which requires precise photon number resolution. Hence, the nonclassicality of the signal is not robust to errors at the detection stage on the order of one photon. One way of visualising this is that for Schr\"{o}dinger cat states, the negativity of the Wigner function is in the interference fringes around the origin (see Fig. \ref{Wigners}.d), which exchange sign for alternating values of $m$. An incoherent mixture of the states heralded by nearby values of $m$, which alternates from even to odd, therefore blurs out these fringes and yields a resultant Wigner function with little overall negativity.

\section{Pure states for intermediate conditioning values}

\begin{figure}
\includegraphics[width=\linewidth]{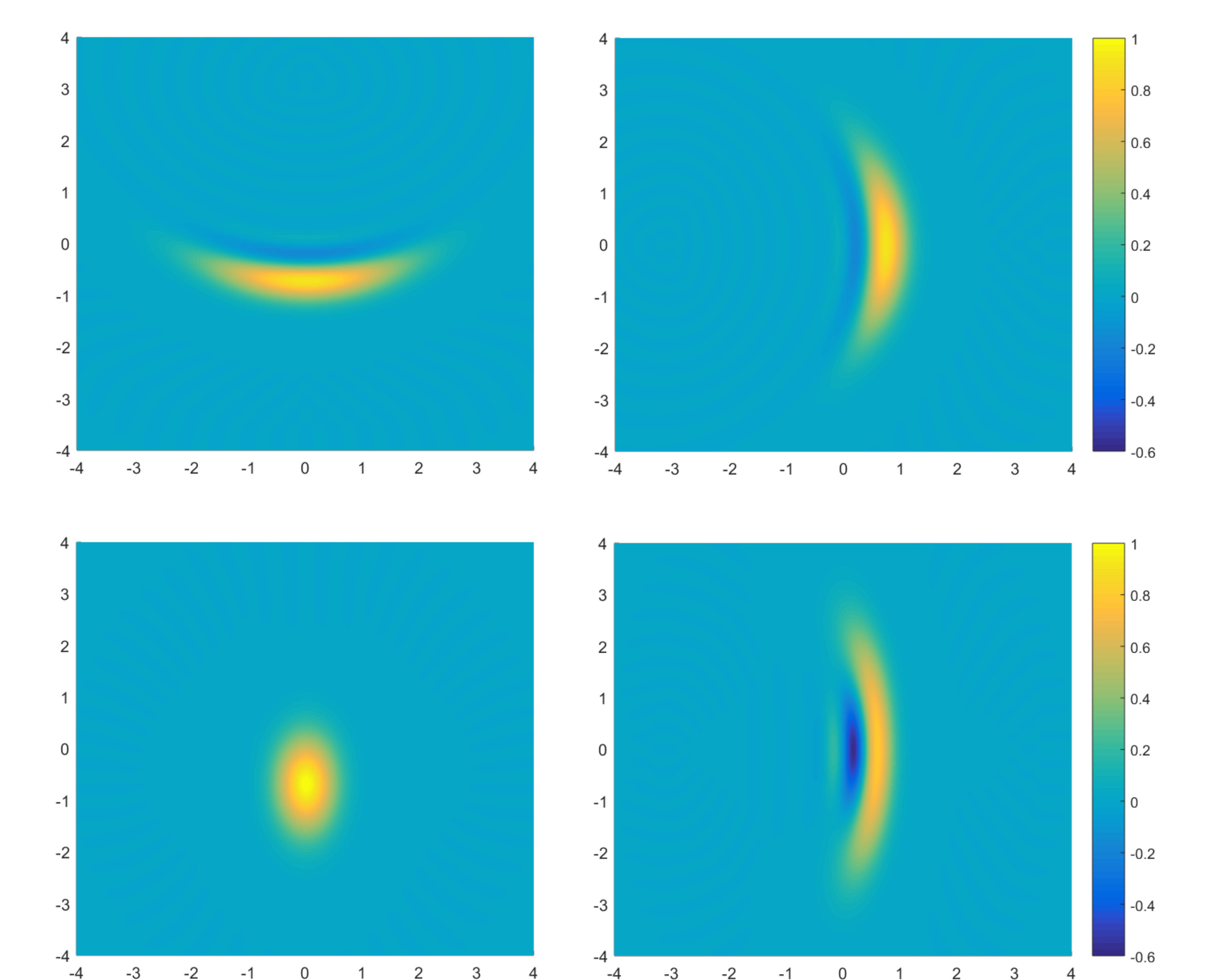}
\caption{Wigner functions for the states heralded by the $m=24$ event at 10 dB squeezing and using (top row) an EPR state and (bottom row) a split squeezed vacuum state. Left column: displacement along the antisqueezed quadrature ($\alpha=4i$). Right column: Displacement along the squeezed quadrature ($\alpha=4$). With the EPR resource, the form of the output state is independent of the displacement field phase and Wigner negativity is weak; with the split squeezed vacuum, Wigner negativity is highly dependent on phase and is strong for real displacement.}
\end{figure}

In this section, we explore the states conditioned by measurement outcomes $m$ in the ancilla that are not much larger than the expectation value $\sim|\alpha|^2$. At this point, the equations can no longer be reduced by approximations and so we must turn to simulations. These were executed numerically on MatLab by explicit calculation of Fock state amplitudes up to 50 photons. To emphasise the aforementioned sensitivity of the scheme to the phase of the displacement field, we also simulated the output states where the input is a two-mode squeezed state, 

\begin{equation}
\Ket{\psi'}_{SA}=\exp\{\xi \hat{a}_S\hat{a}_A-\xi^*\hat{a}^\dagger_S\hat{a}^\dagger_A\}\Ket{\text{vac}}.
\end{equation}

For this state \cite{lvovsky2015squeezed}, which exhibits perfect photon number correlations, the reduced state of both the signal and ancilla modes are thermal states with no well-defined local phase. As such the phase of the displacement field has no physical bearing on the photon statistics of the output $\Ket{\psi'}_{S}$. 

As with the weak-displacement limit, we observe that displacing the ancilla along the squeezed quadrature yields states in the signal mode with strong non-Gaussianity, whereas displacing along the antisqueezed quadrature conditionally prepares very nearly Gaussian states. This contrasts with the EPR state case where, as expected, there is no dependence on the phase of $\alpha$. The postselected state also has weaker negativity than the one prepared by split-squeezed vacuum displaced along the squeezed quadrature. Hence, the maximum non-Gaussianity obtainable from such a resource is less than in the split single-mode squeezed vacuum case.

\begin{figure}
\includegraphics[width=\linewidth]{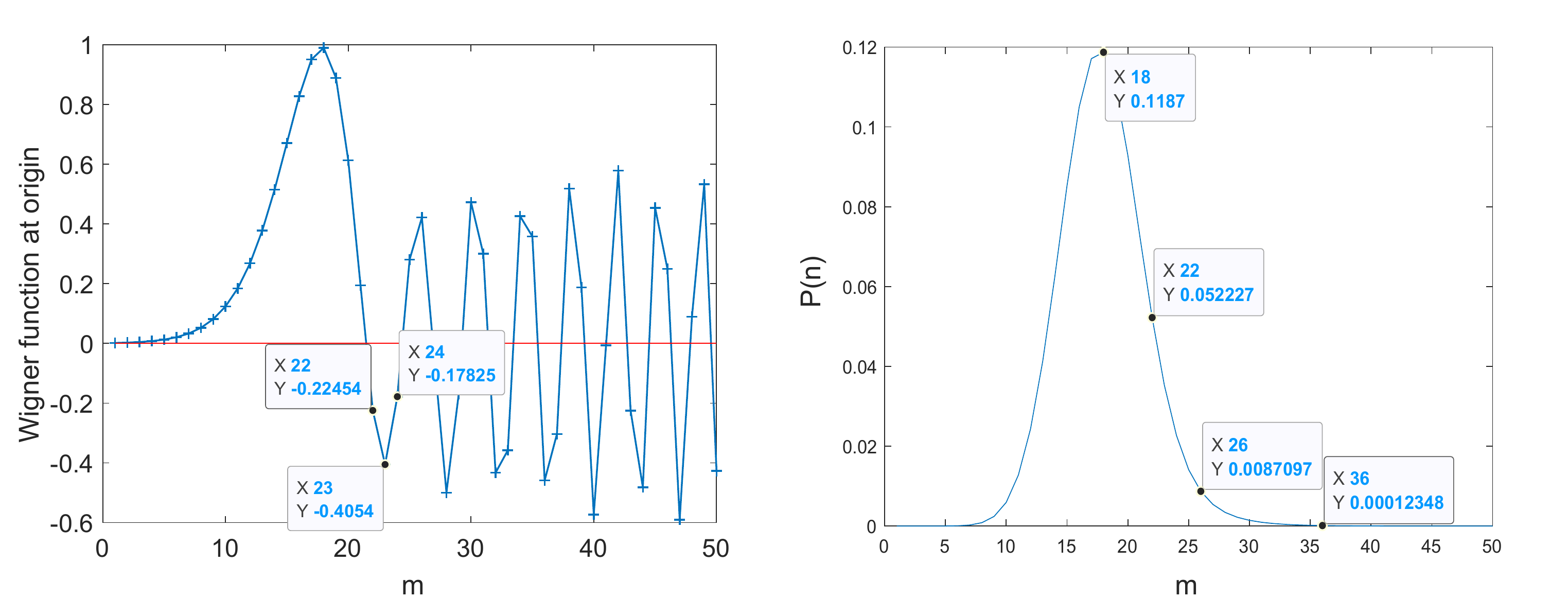}
\caption{Left: Expectation value of the parity operator (equivalent to the value of the Wigner function at the origin) for the states conditioned for various detector outcomes with 10 dB squeezing and $\alpha=4$. Right: Probability distribution for selected detection outcomes in range 1 to 50 for $\alpha=4$ and $10$ dB squeezing.}
\label{parityfig}
\end{figure}

Unlike in the large-$m$ limit, for the events where $m \sim |\alpha|^2$ with real $\alpha$ the parity of the states is no longer so strictly bound to that of $m$. Wigner functions of states heralded by consecutive values of $m$ generally have significant overlaps in their regions of negativity. Fig. \ref{parityfig} shows how the value of the Wigner function at the origin changes as a function of $m$ for the states generated with these parameters. After an initial peak close to the expectation value of $m\approx16$, the parity enters an oscillatory regime, taking both positive and negative values. As already established, at high $m$ the period of these oscillations tends towards 2, as the parity takes alternating positive and negative values. However, at intermediate $m$, the oscillations are slower, and multiple consecutive states have parity of the same sign. A representative example is the three states heralded by $m= \{22,23,24\}$ for the case where the original squeezing is 10 dB and $\alpha$ is 4. The probabilities of heralding these states are 5.2\%, 3.5\% and 2.3\% respectively.  The expectation value of the parity operator for the three states is -0.225, -0.405 and -0.178 respectively. This quantity is equal to the value of the Wigner function at the origin (note that the origin is not the point of maximimum negativity). Since the parity is negative for all three states, all three (and any incoherent mixture thereof) has a Wigner function with a negative region around the origin.  Incoherent mixtures of the states heralded by adjacent values of $m$ (resulting from imperfect photon counting) therefore do not lose their negativity. This allows the heralding of states with deep Wigner negativity even accounting for the non-ideal properties of realistic detectors at mesoscopic light levels, and it is this regime which we contend is the most promising for non-Gaussian state preparation with MSDs.


\begin{figure}[!]
\includegraphics[width=\linewidth]{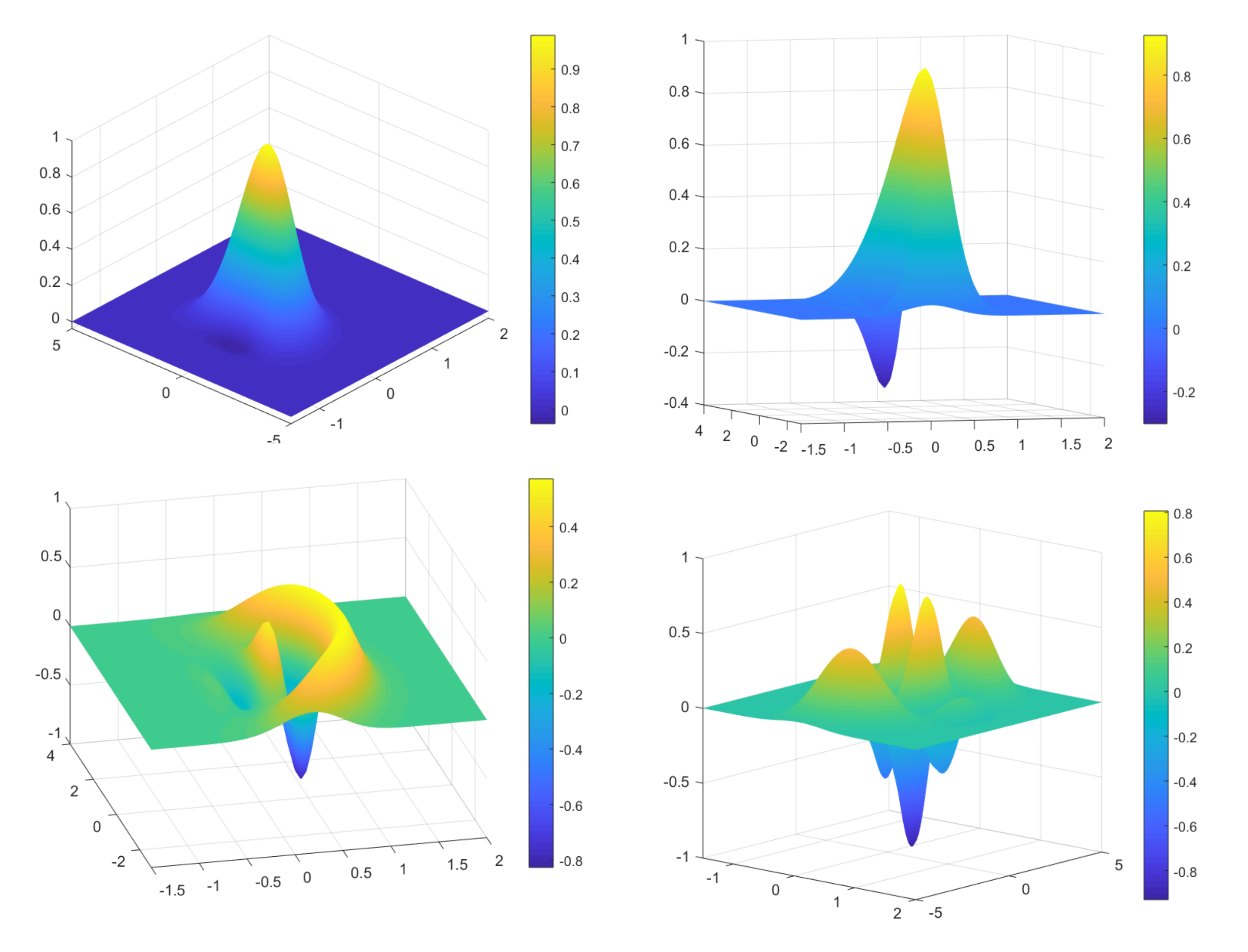}
\caption{Ideal-case Wigner functions for various conditioning outcomes, assuming initial squeezing of $\xi=10$ dB and a displacement parameter $\alpha=4$. From top left: conditioning on 18 photons, 22 photons, 26 photons and 36 photons detected. Note increasing negativity and convergence towards a Schr\"{o}dinger cat state.}
\label{Wigners}
\end{figure}

Fig.\ref{Wigners} shows the conditionally prepared Wigner functions for various measurement outcomes with 10 dB squeezing and $\alpha=4$. Note increasing structure and negativity as $m$ increases, with the form of the states tending increasingly to a Schr\"{o}dinger cat state at high $m$. Fig. \ref{parityfig} (right) shows the probability of heralding these states with ideal conditioning detection.

\section{Modelling experimentally realistic detectors}
Unfortunately, real-world detectors do not exhibit ideal properties and hence cannot be modelled by a projective measurement of the form of Eq. \ref{projective}. In practice, there is a loss of information in the detection and a given detector output $V$ leaves the observer with a mixed state. This can be modelled by a POVM whose elements do not correspond to pure-state operators. The main contributions to this loss of information are non-unit quantum efficiency, the excess noise introduced during signal amplification, and dark current. The effect of all three sources of noise are that the states heralded by certain detector outcomes are incoherent mixtures of the pure states heralded by several values of $m$, with the largest contributions coming from some states with nearby values of $m$.

We assume that the displacement can be performed unitarily, and that the detector itself is stationary, i.e. its response is insensitive to the phase between different Fock-state components in the displaced ancilla. The POVM is therefore diagonal in the displaced Fock basis:

\begin{equation}
\hat{\Pi}_A^V= \sum_m P(V|m)\hat{D}^\dagger_A(\alpha)\Ket{m}_A\Bra{m}\hat{D}_A(\alpha).
\end{equation}

The conditioned state is then given by
\begin{align}
\hat{\rho}_S^V &=N\mbox{Tr}_A\{\hat{\rho}_{SA}\hat{\Pi}_A^V\} \\
&=N\sum_m P(V|m)P(m)\Ket{\psi_m}_S\Bra{\psi_m}, \label{prebayes}
\end{align}
where $P(m)$ is the probability of a displaced $m$-photon event such that
\begin{equation}
P(m)=\mbox{Tr}_S\{\Bra{\psi}_{SA}\hat{D}^\dagger_A(\alpha)\Ket{m}_A\Bra{m}\hat{D}_A(\alpha)\Ket{\psi}_{SA}\}.
\end{equation}
To preserve the unity of the trace the normalisation constant $N$ is given by $N=1/P(V)$, i.e. just the inverse of the probability $P(V)$ of recording the $V$-detection event, where
\begin{equation}
P(V)=\sum_p P(V|m)P(m).
\end{equation}
Substituting this into Eq. \ref{prebayes} and using Bayes' theorem:
\begin{equation}
P(m|V)=\frac{P(V|m)P(m)}{P(V)},
\label{Bayes}
\end{equation}
the normalised conditioned state is given by
\begin{equation}
\hat{\rho}_S^V=\sum_m P(m|V)\Ket{\psi_m}_S\Bra{\psi_m},
\end{equation}

where $P(m|V)$ is a relative weight for the classical probability of having the displaced $m$-photon Fock state given the $V$-detection event, which can be calculated from the detector response $P(V|m)$ via. Eq. \ref{Bayes}.

We first model the non-unit quantum efficiency of the detector $\varepsilon \leq 1$. In this case, the possible detector outcomes $V$ correspond to events where only $b$ photons are absorbed by the active part of the detector from the $m$ photons in the signal, with $b\leq m$. Since the probability of each photon being absorbed by the detector is independent, the probability of $b$ absorption events given a signal state containing $m$ photons is modelled by the binomial distribution
\begin{equation}
P(b|m)=\frac{m!}{b!(m-b)!}\varepsilon^b (1-\varepsilon)^{m-b}.
\end{equation}
For an amplified photodetector such as an avalanche photodiode, another contribution to the mixedness is the noise due to the amplification process, quantified by the  excess noise factor \cite{Teich1986}. A single absorption event gives rise to an avalanche of $M$ daughter carriers, where $M$ is the gain random variable. Assuming that the contribution to the photocurrent from each carrier-amplification event is independent, due to the central limit theorem in the limit of large $b$ the statistics of the overall photocurrent are determined entirely by the mean $\langle M \rangle$ and variance $\mbox{Var}(M)=\langle M^2 \rangle - \langle M \rangle^2$. The variance may be related to the excess noise factor, defined as the normalised second moment of the gain $M$,
\begin{align}
F_e\equiv& \langle M^2 \rangle/\langle M \rangle^2 \\
=& \mbox{Var}(M)/\langle M \rangle^2+1.
\end{align}

In the case of $b$ absorption events, and assuming that the $b$ multiplication processes are independent, then the total number of electrons output from the detector is $V$, where $\langle V \rangle=b\langle M \rangle$ and the variance of $V$ given $b$ absorption events is given by $\mbox{Var}(V)=b\mbox{Var}(M)$. The probability distribution of $V$ is then given by
\begin{align}
P(V|b)=&\exp\left( \frac{-(V-\langle V\rangle)^2}{\mbox{Var}(V)}\right) \\
=&\exp\left( \frac{-(V/\langle M\rangle-b)^2}{b(F_e-1)}\right)
\end{align}

The resultant detector response to an $m$-photon event is then given by 
\begin{equation}
P(V|m)=\sum_b P(V|b)P(b|m).
\end{equation}

The effect of dark current, or baseline electronic noise, is straightforwardly modelled by a convolution of the signal with the spread function due to noise. This is strongly application-dependent, and for the applications using pulsed light and detectors with greater bandwidth than the repetition rate, the dark current can largely be eliminated by temporal gating. 

\begin{figure}[!]
\includegraphics[width=\linewidth]{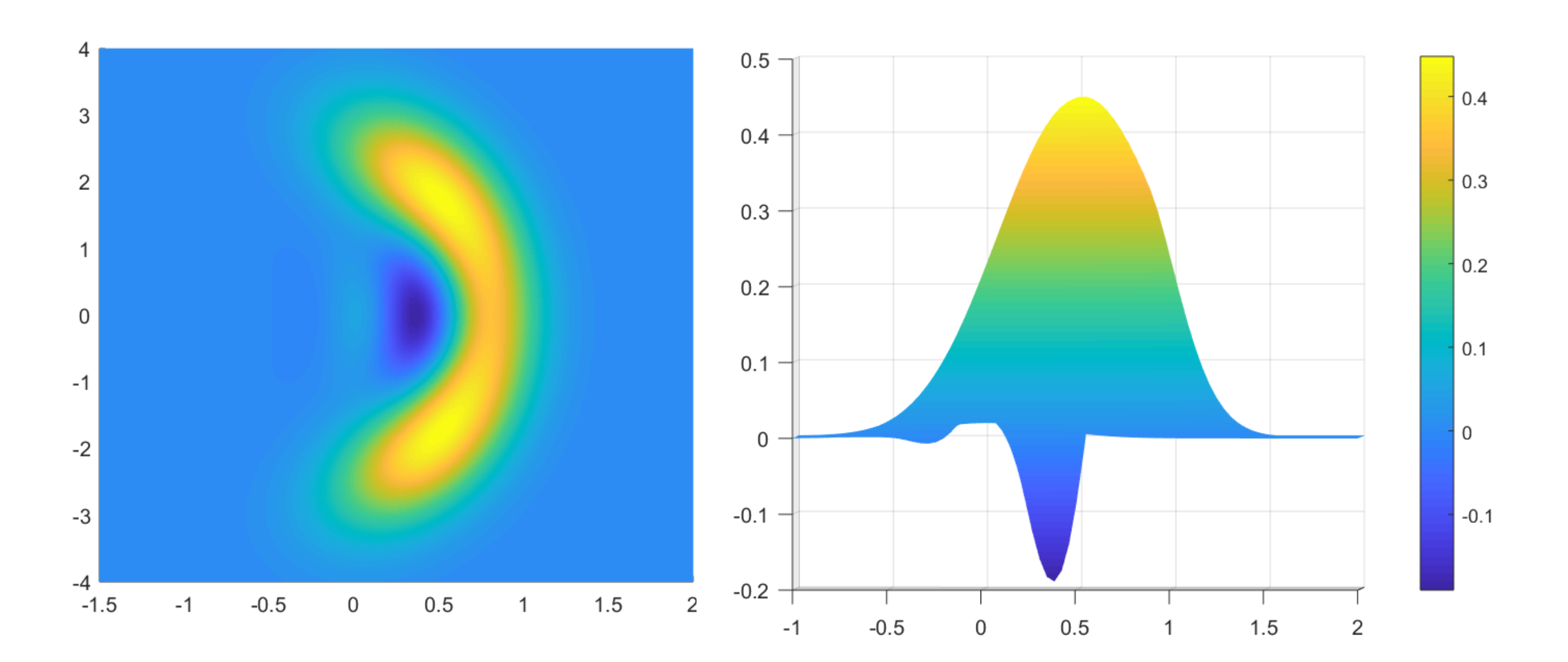}
\caption{Wigner function for a state heralded by imperfect mesoscopic detection with 10 dB squeezing, $\alpha=4$ and modal conditioning number $m=25$. Quantum efficiency is 90\% and $F_e=1.1$.}
\label{LossyWigner}
\end{figure}

Fig. \ref{LossyWigner} shows the result of these non-idealities for a state produced by imperfect mesoscopic detection with 10 dB initial squeezing, $\alpha=4$ and photon number $m\approx25$. The quantum efficiency of the detection is taken to be 90\% and excess noise $F_e=1.1$, which are values that are experimentally plausible at the current state of the art.

\section{Experimental feasibility}

The 10 dB squeezing level used in the simulations presented is already relatively conservative by cutting-edge experimental standards. Experiments using optical parametric oscillators (OPOs) have demonstrated AC squeezing of 12.3 dB at 1550 nm \cite{Mehmet2011} and 15 dB at 1064 nm \cite{Vahlbruch2016}. Some promising MSDs, such as HgCdTe detectors \cite{vojetta2012linear, Dumas2017}, have good broadband response across the infrared region and so represent a promising candidate system for conditional state preparation in conjunction with such OPOs. 

In practice, many sources of squeezed vacuum are multimode, whether spectrally such as a synchronously-pumped OPO \cite{Arau2012}, or spatially like many single-pass PDC sources. This multimode nature provides both challenges and opportunities. On the one hand, the presence of multiple squeezed modes allows for the automatic creation of non-Gaussian entanglement by performing a non-Gaussian measurement on a superposition of eigenmodes of the coviariance matrix. On the other hand, in the multimode case photon counting without mode resolution incurs a loss of information which can severely degrade the purity of the state. In previous demonstrations with conditional preparation with multimode Gaussian resources, such as photon subtraction, experimentalists have employed a quantum pulse gate to ensure (post-selectively) that the measurement is only performed on a single selected mode \cite{Ra2019}. However, this is often an experimentally taxing process, requiring efficient optical nonlinearities and synchronisation with an independent optical gate field. In our method, mode selectivity is achieved by the choice of mode of the displacement field. Provided that the photon number occupation of the other modes is small compared to $|\alpha|^2$, their overall contribution to the heralding value $m$ will be small and the statistics of the conditioning measurement will be dominated by only the desired mode. This intrinsic mode-selectivity is a significant experimental advantage of our scheme.

A major appeal of using MSDs is the higher repetition rate (or measurement bandwidth) they are capable of sustaining. Whereas commercially available SNSPD systems are typically specified with maximum count rates of a few megahertz, the bandwidths of mesoscopic APDs can be orders of magnitude higher. In conjunction with their higher saturation threshold, this means that MSDs can monitor optical fields with squeezing up to tens or hundreds of megahertz (consistent with table-top oscillators) without loss of quantum information, and therefore have the potential to yield far higher generation rates than attainable with existing sources of heralded single photons.  

\section{Conclusions and outlook}
Here we have shown how non-Gaussian states with negative Wigner functions can be generated from a single-mode squeezed vacuum resource state using only displacement and photon-number resolved measurement in the mesoscopic regime. This protocol can be used to generate Schr\"{o}dinger cat states in the limiting case, for high-intensity conditioning events with weak displacement. For non-negligible displacements, the phase of the coherent displacement field relative to that of the local squeezing is extremely important to the quantum properties of the conditioned state, with displacement along the noisier quadrature conditioning states that are approximately coherent, and displacement along the more squeezed quadrature generating states with high non-Gaussianity. However, in this limit, the negative regions of the Wigner functions of the heralded states are non-overlapping and hence the Wigner negativity is not robust to imprecise photon number resolution in the conditioning detector.

At intermediate photon numbers in the ancilla mode, the different signal states prepared for several consecutive conditioning measurement outcomes have Wigner functions whose negative regions overlap in phase space, and hence an incoherent mixture of them (as occurs with imprecise photon number resolution) still has negativity. Simulations show that for a realistic model of these detectors and with attainable squeezing values, it is possible to prepare states with significant negativity, with rates of tens to hundreds of megahertz.

 Simulations with a pure two-mode squeezed state (also known as an Einstein-Podolsky-Rosen or EPR state) show that the nonclassicality of the conditioned state rapidly deteriorates with increasing displacement $|\alpha|$, emphasising the importance of the local squeezing on the state for retention of quantum properties up into the mesoscopic regime. Future work may nonetheless wish to consider the more general form of two-mode Gaussian entanglement, with varying degrees of photon-number correlation and local squeezing on the signal and ancilla. Additionally, generalisations of this strategy to a multimode picture for quantum information applications present an exciting line of enquiry.

Experimentally, we have considered the effect of realistic loss and noise on the quality of the states prepared for feasible squeezing levels. The mode-selective character of the conditioning could allow the heralding of nonclassical states from a multimode source or even the preparation of non-Gaussian entanglement between separate modes, all without the need for optical nonlinearities (notwithstanding the initial squeezed state preparation). Additionally, the high bandwidths of MSDs on the near horizon promises much higher preparation rates than can be achieved natively with a leading single-photon counters such as SNSPDs. 

This work received funding from the European Union's Horizon 2020 research and innovation programme under grant agreement No 899587. This work was supported by the European Union's Horizon 2020 research and innovation programme under the QuantERA programme through the project ApresSF and  by the European Research Council under the Consolidator Grant COQCOoN (Grant No. 820079).

\section{Appendices}
\subsection{Preparation of Schr\"{o}dinger cat states}
\label{focklimit}

We consider the limit where $|\alpha|^2=0$ and hence $\Delta_{mk} = \delta_{mk}$. The first observation to make about this case is that the parity of $\Ket{\psi_m}_S$ is the same as that of the heralding count $m$. Specifically,
\begin{align}
\Ket{\psi_{m \text{ even}}}_S=& N_m\sum_{j=0}^\infty c_{j+\frac{m}{2}} \frac{(2j+m)!}{\sqrt{m!}} \frac{\Ket{2j}_S}{\sqrt{(2j)!}} \label{even}
\end{align}
for $m$ even, and 
\begin{align}
\Ket{\psi_{m\text{ odd}}}_S=& N_m\sum_{j=0}^\infty c_{j+\frac{m+1}{2}} \frac{(2j+m+1)!}{\sqrt{m!}} \frac{\Ket{2j+1}_S}{\sqrt{(2j+1)!} }
\end{align}
for $m$ odd. We can write the coefficients as 
\begin{align}
&c_{\mu} \frac{(2\mu+2j)!}{\sqrt{m!}}\\
= &\frac{(-1)^{\mu+j}}{4^{\mu+j}\sqrt{\cosh r}} \frac{(2\mu+2j)!}{(\mu+j)!}\frac{(\tanh r)^{\mu+j}}{\sqrt{m!}} \label{bitch}
\end{align}
for $\mu=\{m/2,(m+1)/2\}$. For the events where $\mu\gg j$, we can use the approximation (which follows from the Stirling approximation)
\begin{equation}
\frac{(2\mu+2j)!}{(\mu+j)!} \approx \frac{\mu!}{\sqrt{\pi \mu}}4^{\mu+j}\mu^j,
\end{equation}
and hence the right hand side in Eq. \ref{bitch} can be written
\begin{align}
\frac{(-1)^{\mu}(\tanh r)^\mu\mu!}{\sqrt{m!\pi \mu\cosh r}} (-\mu\tanh r)^j.
\end{align}
We can therefore write Eq. \ref{even} in the form
\begin{align}
\Ket{\psi_{m \text{ even}}}_S=& N'_m\sum_{j=0}^\infty \beta^{2j} \frac{\Ket{2j}_S}{\sqrt{(2j)!}} \\
=&\frac{\Ket{\beta}+\Ket{-\beta}}{\sqrt{2}},
\end{align}
where $\Ket{\beta}$ is a coherent state with $\beta=i\sqrt{m\tanh r/2}$, and similarly for $m$ odd (with $\beta'=i\sqrt{(m+1)\tanh r/2}$)
\begin{align}
\Ket{\psi_{m \text{ odd}}}_S=\frac{\Ket{\beta'}-\Ket{-\beta'}}{\sqrt{2}}.
\end{align}
This recovers the result found originally in \cite{Dakna97}.

\bibliography{BiblioMeso}

\end{document}